# Empirical Exploration of Deformation Mechanisms in Deep Rock Exposed to Multiple Stresses

Ali Nassiri[1]*
[1] University degli studi di milano, Milano, Italy.
*Corresponding author: Ali.nassiri@studenti.unimi.it

**ABSTRACT**
To avoid the impact of inherent natural imperfections on experimental outcomes during testing, a recently designed genuine triaxial apparatus has enabled the replication of conditions where the three principal stresses exhibit varying magnitudes. This setup facilitates the examination of fracture behaviors in materials similar to the 4-series under genuine triaxial stress states. The experimental findings revealed that the maximum strength experiences significant influence from the minimum principal stress and the intermediate principal stress in genuine triaxial stress conditions. Specifically, for a fixed minimum principal stress ($\sigma_3$), there is a progressive reduction in peak strain as the intermediate principal stress ($\sigma_2$) increases. Moreover, as the minimum principal stress ($\sigma_3$) rises, the magnitude of peak strain diminishes. The Young's modulus of the specimens exhibits an exponential increase with the augmentation of confining pressure, and it demonstrates a linear growth pattern as the intermediate principal stress increases. Additionally, the cohesion force is observed to rise while the internal friction angle decreases with the elevation of the intermediate principal stress. The failure modes of the specimens in genuine triaxial stress conditions can be categorized into three types, all of which exhibit a specific angle between the fracture surface and $\sigma_3$, ranging from 50 to 62 degrees. Furthermore, a comparison between the computed range and the actual measurement range of the fracture angle highlights several limitations in the relationship between the fracture angle ($\theta$) and the internal friction angle ($\varphi$) within the Coulomb criterion, particularly in rocks containing complex fracture systems.

**INTRODUCTION**
The exploration and exploitation of oil and gas resources have increasingly ventured into deeper territories due to rising demands and the continuous extraction from shallow strata. However, existing methodologies face growing challenges and reliability issues when applied in deeper, hotter, and more complex geological formations. Consequently, the investigation of deformation and failure mechanisms in deep-seated rock under complex stress conditions using a newly developed true triaxial machine [1–3] holds significant theoretical and practical importance. It's widely acknowledged that the deformation and fracture behaviors of subsurface rock under complex stress states are intricate, compounded by the current lack of precise and effective monitoring techniques [4–6]. Directly observing the mechanical characteristics and failure modes of deep-seated rocks in the field is often impractical, necessitating laboratory true-triaxial tests. Mahmoudabadi's research underscores the importance of viscous dampers in fortifying the seismic resilience of reinforced concrete frames. Additionally, a study on cable behavior, including the influence of spring dampers and viscous dampers under dynamic loads, offers crucial insights into structural dynamics. Further exploration by Mahmoudabadi delves into enhancing suspension bridges' structural performance, emphasizing the valuable role of viscous dampers in structural engineering. [7–9]

Unlike conventional triaxial equipment, which can only apply two different principal stress magnitudes, true triaxial apparatus can generally be categorized into three types based on the stress mode applied to the rock's surface [10–14]. Sweden pioneered the world's first true triaxial apparatus [15], while Mogi et al. [16] independently developed a true triaxial apparatus that addressed the sample alignment challenge in true triaxial loading systems. Additionally, they used this apparatus to comprehensively examine the relationship between compressive strength and the intermediate principal stress of rocks under true triaxial stress [17,18]. The advancement of true triaxial apparatus has significantly enhanced the progress of true triaxial testing. Xu et al. [19,20] investigated the relationship between the ratio of intermediate principal stress and rock strength and deformation using true triaxial apparatus. Yin et al. [21,22] conducted indoor true triaxial tests on sandstone and limestone, comparing the results of these two types of rock under different stress paths. Yang et al. [23] performed true triaxial tests on rock strength using a true triaxial apparatus, discovering that as the intermediate principal stress increased, so did the peak strength of the rock. As the intermediate principal stress increased from lower values, the rock transitioned from plasticity to brittleness. Li et al. [24] conducted an extensive series of sandstone true triaxial loading tests utilizing a self-developed true triaxial electro-hydraulic servo control system. Combined with CT scanning technology, they systematically studied the strength, deformation, and fracture characteristics of sandstone under varying intermediate principal stress conditions.

Through this comparison, it becomes evident that both domestic and international experts have developed various types of true triaxial apparatus. However, there is a significant gap in the methodology and scope of research in true triaxial testing, depending on different subjects and objectives [25]. As our understanding of rock mechanics deepens, it is increasingly recognized that rock mechanical properties are closely linked to the breaking characteristics and intermediate principal stresses of rocks. Thus, there is a pressing need to investigate the strength, deformation, and failure patterns of rocks under true triaxial stress states [26–29]. Shabbir et al. offer a comprehensive exploration of construction defects in reinforced concrete corbels, combining experimental and numerical analyses to enhance our understanding of structural integrity. [30]

To bridge the gap between laboratory test results and field geological work, some argue that larger specimens are more representative, making experimental results closer to real-world conditions. However, obtaining large-sized specimens is challenging due to the high cost of natural core sampling, processing difficulties, and limitations in drilling technology. Additionally, the heterogeneity of natural rocks, stemming from sedimentary and structural factors, may also influence experimental outcomes. Therefore, we initially employ a similar material physical simulation test method, using a true triaxial machine to examine the strength, deformation, and failure characteristics of rocks under different in-situ stress conditions.

**MATERIALS AND METHODS**

In this research, all experiments were conducted using the YSZS-2000 type multi-functional electric-hydraulic servo-controlled rock true triaxial machine, as depicted in Figure 1. This machine was designed and manufactured in collaboration between Northwest University and Changsha Ya-xing CNC Technology Limited Company. Its primary components consist of the main framework, electro-hydraulic servo loading system, lifting mechanism, electro-hydraulic servo supercharging device, measurement and control system, and computer control and processing system. The hydraulic servo system, at the core of this machine, is capable of regulating the deformation speed based on rock failure and deformation criteria while maintaining a constant deformation speed. Additionally, the testing machine requires substantial stiffness to minimize the accumulation of strain energy within it. This machine's stiffness exceeds 10 GN/m, which adequately satisfies the demands of our tests.

The equipment can replicate 3D stress conditions by controlling the magnitudes and orientations of the maximum, intermediate, and minimum principal stresses within 100×100×100 mm³ specimens. The maximum stress and intermediate principal stress are applied using rigid platens through two pairs of orthogonal electro-hydraulic servo loading systems. Pressure magnitude is measured using a Hub-and-Spoke load pressure sensor. The minimum principal stress is applied through a soft loading system by confining fluid pressure inside the pressure chamber, and this pressure is monitored using a hydraulic pressure sensor. The servo-controlled system has a maximum loading capacity of 2000 KN in the σ1 and σ2 directions, with a maximum confining pressure of 100 MPa in the σ3 direction. Throughout the experiments, axial and lateral deformations are measured using two pairs of linear variable displacement transducers (LVDTs) with a range of approximately 80mm. All three principal stresses are monitored and controlled through a closed-loop servo control system.

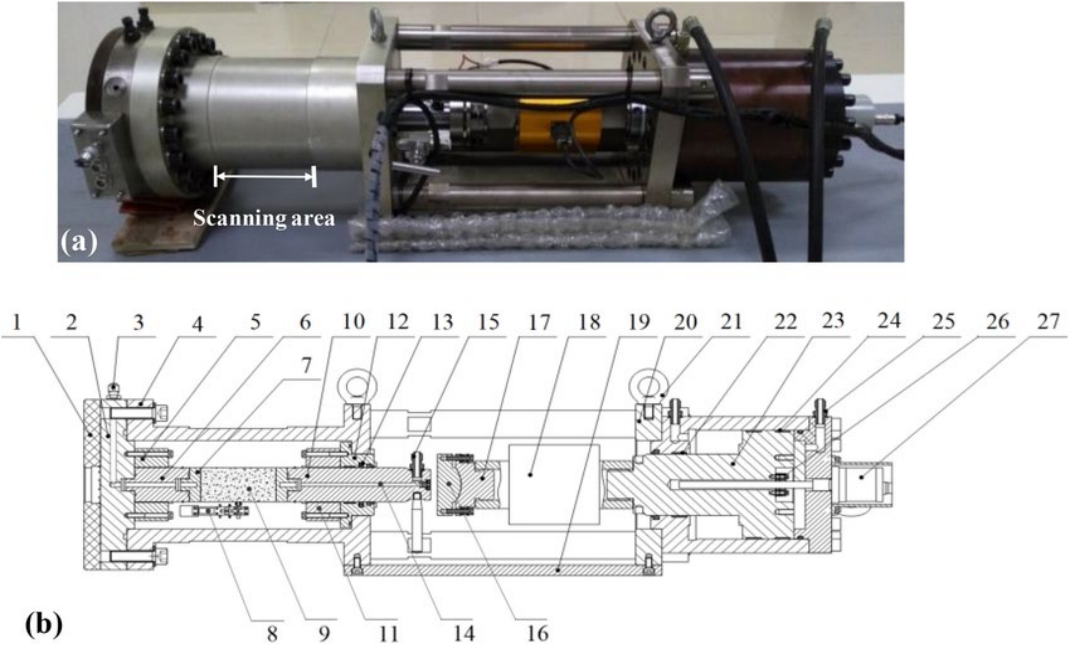

**FIGURE 1. Structure of true tri-axial pressure chamber**

Similar materials typically consist of cementation and filler, where the mechanical properties of the cementation play a pivotal role in determining the overall mechanical characteristics of the similar materials. When selecting materials for similar materials, the following principles are generally considered, based on previous experience:

Granular aggregate materials are preferred as they are easy to compact and prepare after the addition of cementing agents.

The selected materials should exhibit stable mechanical properties regardless of temperature and humidity variations.

Materials should be readily available in quantity and have a reasonable cost.

Only non-toxic materials that are safe for both humans and the environment should be used.

Tests have shown that using cement with higher strength allows for a broader range of mechanical adjustments through ratio changes, making it a widely employed choice [31–33]. Sand, often used as a filler in similar materials, significantly influences the strength of the material, with damage typically occurring between sand particles.

For this particular experiment, the materials used are as follows: The cementation consists of P.C32.5R composite Portland cement produced by Yuncheng, Shanxi. The sand selected for this experiment was filtered using a square hole sieve with a 0.9mm mesh size.

The preparation process involves molding the similar materials naturally [34]. Standard specimens are created using specialized molds (100×100×100 mm³ and φ50×100mm), as illustrated in Figure 2. The steps for specimen preparation are as follows:

Mix the aggregate and cementing material according to the specified proportions. Add the required amount of water and stir the mixture with a muddler until lumps dissolve and air bubbles are expelled. Pour the resulting liquid into the special molds to form cubic specimens. Allow the samples to set for a certain period, and then remove them from the molds.

Before conducting tests, dry the specimens in an oven at 40°C for a minimum of 24 hours.

After complete drying, weigh the specimens to determine their mass, and calculate the density accordingly.

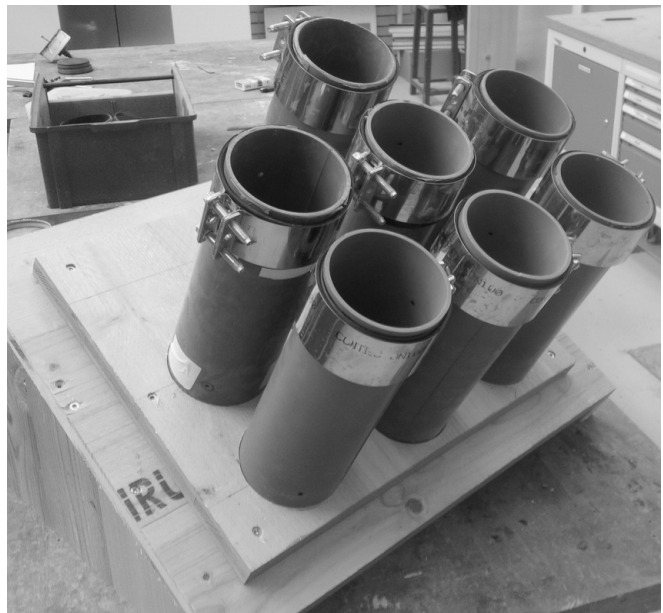

**FIGURE 2**
**The special molds and a part of specimens**

## RESULTS AND DISCUSSION

When studying the deformation characteristics of cement specimens under true triaxial compression conditions, previous research has explored the relationship between confining pressure and elasticity modulus. It has been observed that experimental results differ significantly across various types of rocks. One of the earliest studies in this regard was conducted on griotte through uniaxial and tri-axial tests. The findings from these tests indicated that the elasticity modulus of griotte remains constant and does not change with variations in confining pressure. In fact, it is equal to the elasticity modulus observed in uniaxial compression tests. Furthermore, these results suggested that confining pressure exerts no significant influence on the elasticity modulus in the case of griotte.

However, it's important to note that the behavior of sedimentary rock differs from griotte. In sedimentary rock, the elasticity modulus increases as the confining pressure increases. The stress-strain curve in rock compression is typically non-linear, and there are generally three methods used to represent it: the tangent modulus, the secant modulus, and the average modulus. The average modulus is determined by calculating the average slope of the relatively straight portion of the stress-strain curve, which reflects the proportional relationship between changes in strain and stress. This method is less affected by experimental conditions and carries specific mechanical significance. Consequently, when studying the mechanical properties of rock, the average modulus is commonly used to characterize rock deformation characteristics. The Young's modulus mentioned in this article refers to the average modulus.

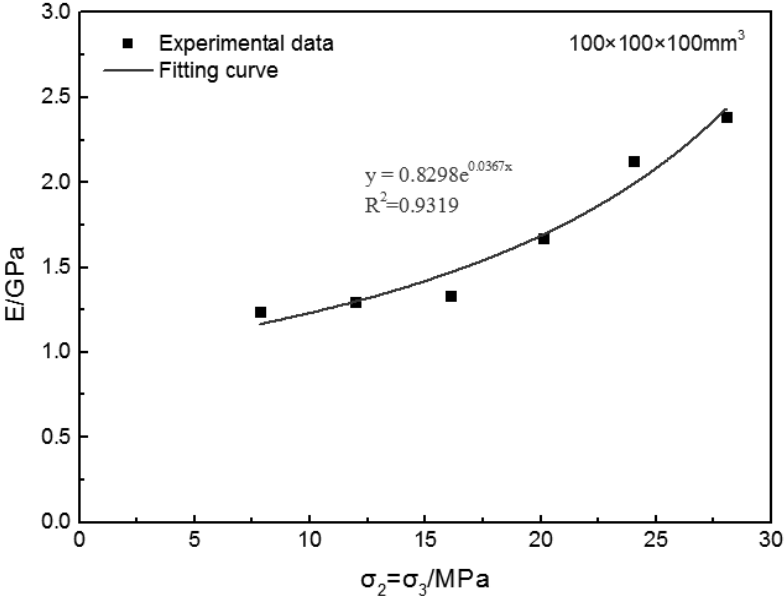

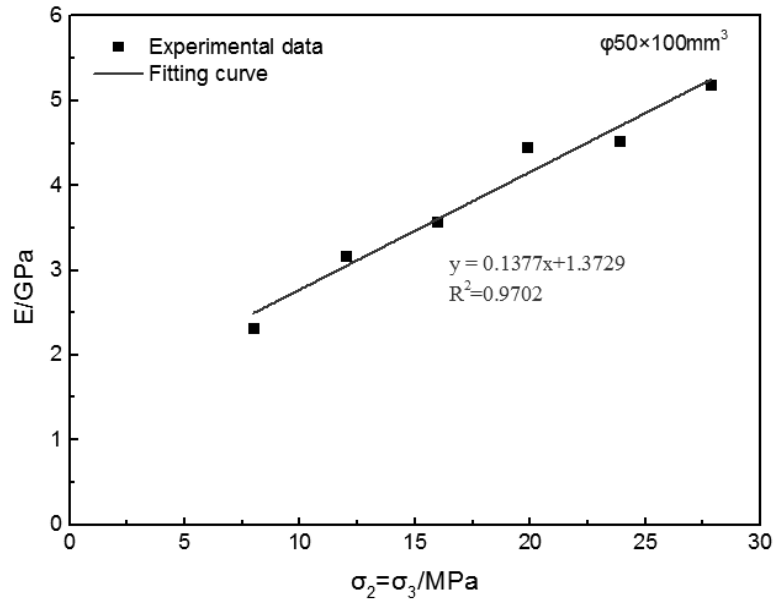

**FIGURE 3. The relationship between Young's modulus and confining pressure**

The relationship between Young's modulus and confining pressure is illustrated in Figure 3, with two scatter plots representing cube and cylindrical specimens. Both figures demonstrate an increase in Young's modulus as confining pressure rises. This behavior occurs because as confining pressure increases, the inherent pores within the specimen tend to close. Additionally, the difference in specimen sizes leads to variations in Young's modulus with changing confining pressure. The increase in Young's modulus for cube specimens with increasing confining pressure is relatively modest, while it becomes more pronounced as the confining pressure increases. This suggests an exponential relationship between Young's modulus and confining pressure for cube specimens. Conversely, the relationship for cylindrical specimens appears to be linear.

Furthermore, the figures provide specific information about Young's modulus values. For cube specimens, Young's modulus falls within the range of 1.21-2.39 GPa under pressures ranging from 8 to 28 MPa, while for cylindrical specimens, it falls within the range of 2.303-5.162 GPa. Notably, the Young's modulus of cube specimens is significantly lower than that of cylindrical specimens under equivalent confining pressures. This difference can be attributed to two main factors:

Specimen size: Generally, as specimen size increases, the strength of the specimens tends to decrease. Therefore, cube specimens exhibit a relatively lower Young's modulus compared to cylindrical specimens.

Specimen shape: The distribution of confining pressure on cylindrical specimens is more uniform than on cube specimens, resulting in higher Young's modulus values for cylindrical specimens. The corners of the cube specimens experience a marginal effect, which contributes to the lower Young's modulus observed.

The relationship between peak strain and intermediate principal stress is a critical aspect explored in this study. Peak strain is defined as the axial strain corresponding to the point where the specimen reaches its maximum strength. In general, brittle failure tends to occur at lower axial strains under uniaxial compression conditions. As horizontal stresses increase, the strength of the specimen also increases. For instance, under uniaxial compression, the specimen exhibits a peak strength of 17.21 MPa and a peak strain of 2.953%.

Figure 5 illustrates that there is a linear negative correlation between the intermediate principal stress and peak strain, considering a fixed σ3. In other words, as the intermediate principal stress increases, the peak strain decreases. However, the rate of decrease in peak strain is influenced by the minimum principal stress. The analysis reveals that the slope of the fitting line gradually decreases as the minimum principal stress increases. This suggests that the decrease in peak strain becomes less pronounced with higher minimum principal stresses. Nevertheless, it's worth noting that the peak strain eventually reaches a value comparable to the peak strain observed under uniaxial compression within the brittle failure range.

This observation underscores that the variation in peak strain is not solely dependent on the intermediate principal stress; it is also influenced by the minimum principal stress. In essence, the variation in peak strain is the outcome of the interaction between the intermediate principal stress and the minimum principal stress. These findings emphasize the complex interplay of multiple stress factors in determining the behavior of peak strain in the tested specimens.

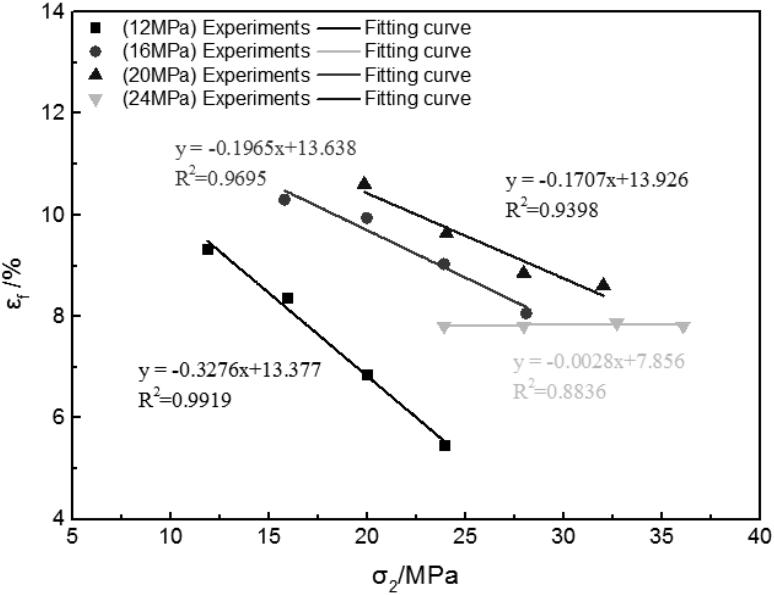

**FIGURE 5. Relationship between peak strain and intermediate principal stress**

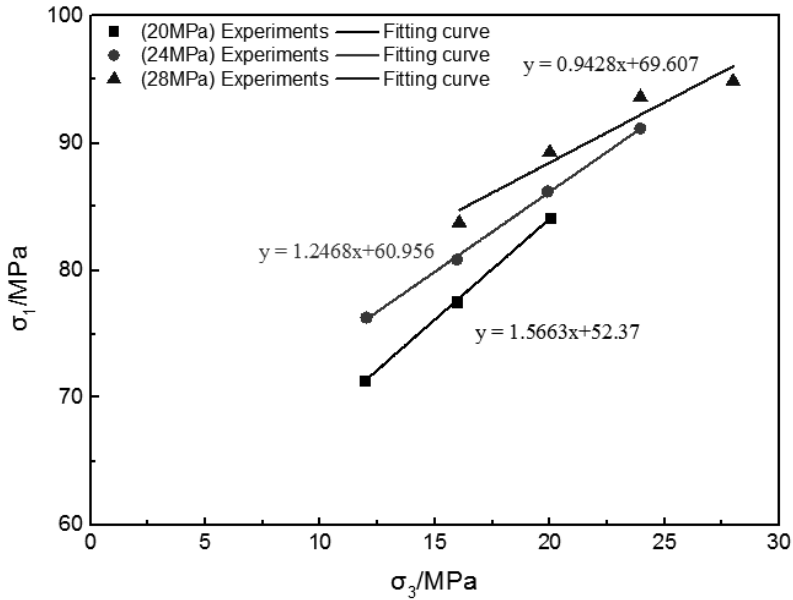

**FIGURE 6. Relationship between peak strength and minimum principal stress**

The strength characteristics of cement specimens under true triaxial compression conditions are explored in this study. Let's examine the key findings:

Relationship Between Peak Strength and Minimum Principal Stress for Cube Specimens (Figure 6): The results indicate that the minimum principal stress significantly impacts the peak strength of cube specimens. Specifically, as the minimum principal stress increases, the peak strength also increases for a given intermediate principal stress. However, the magnitude of the increase in peak strength is notably smaller as the intermediate principal stress increases. In summary, minimum principal stress has a pronounced effect on peak strength, and this effect is modulated by the intermediate principal stress.

Influence of Intermediate Principal Stress on Peak Strength (Table 1 and Figure 8): The peak strength of specimens in different stress states is presented in Table 1. The data show that peak strength increases with an increase in the intermediate principal stress while keeping the minimum principal stress constant. Figure 8 illustrates the relationship between intermediate principal stress and peak strength for various minimum principal stress levels. It is evident that peak strength rises as the intermediate principal stress increases. When comparing the slopes of the fitting lines for different stress conditions, it becomes apparent that the slope of the fitting line decreases as the minimum principal stress increases. This suggests that the influence of intermediate principal stress on peak strength diminishes as the minimum principal stress rises. In essence, when the minimum principal stress is lower, increasing the intermediate principal stress enhances the specimen's compactness, leading to a significant increase in peak strength. Conversely, when the minimum principal stress is higher, it already contributes to specimen compaction, reducing the impact of intermediate principal stress on peak strength.

These findings highlight the complex interplay of minimum and intermediate principal stresses in determining the peak strength of the tested specimens. The relationship between these stress parameters is characterized by intricate dependencies that vary with different stress conditions, demonstrating the nuanced behavior of the cement specimens under true triaxial compression conditions.

The analysis from Figure 7 suggests that peak strength is indeed influenced by both the minimum principal stress and intermediate principal stress. However, it's evident that the impact of the intermediate principal stress on peak strength is notably smaller compared to that of the minimum principal stress. This observation can potentially be attributed to the different loading modes associated with the intermediate and minimum principal stresses.

In the context of this analysis, the direction of σ2 represents a rigid loading mode, while the direction of σ3 corresponds to a flexible loading mode. The key distinction here lies in the nature of the loading: flexible loading distributes force more uniformly than rigid loading. As a result, when the minimum principal stress (σ3) is increased for a fixed σ2, the increase in peak strength is more pronounced compared to when σ2 is increased for a fixed σ3.

In simpler terms, increasing σ3 in a flexible loading mode has a more substantial effect on peak strength when compared to increasing σ2 in a rigid loading mode. This observation aligns with the notion that the loading mode can significantly influence the behavior of the tested specimens under true triaxial compression conditions, highlighting the importance of considering both minimum and intermediate principal stresses in analyzing peak strength.

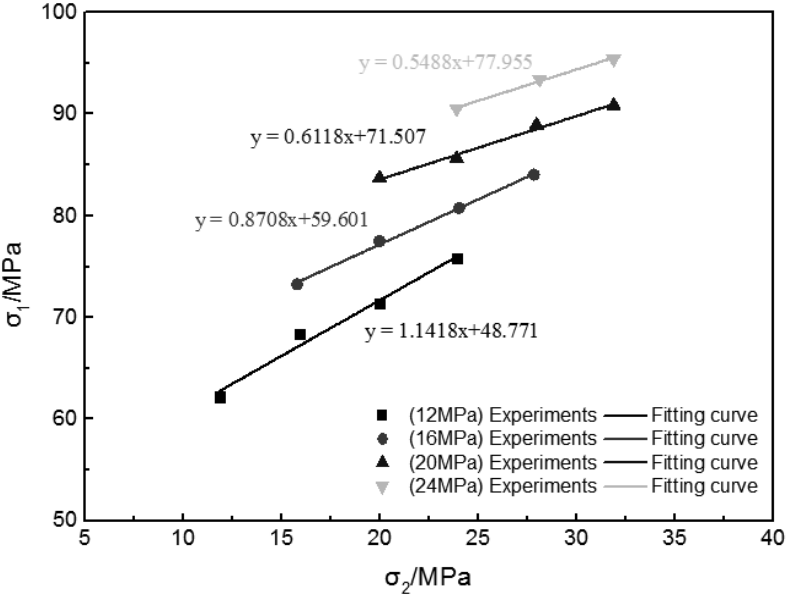

**FIGURE 7. Relationships of peak strength and intermediate principal stress**

The influence of the intermediate principal stress on the shear strength index is examined in this study. The process of handling true triaxial data involves constructing Mohr stress circles in the τ-σ stress plane, with the normal stress as the transverse coordinate and the shear stress as the vertical coordinate. The center of these circles is ((σ1+σ3)/2, 0), and the radius is (σ1-σ3)/2. By changing the minimum principal stress, researchers draw 3-4 Mohr stress circles for the same intermediate principal stress and then determine a common tangent to these circles, which results in a stress Mohr envelope curve. In the τ-σ coordinate system, the cohesive force (c) is determined as the intercept of the envelope curve with the τ axis, while the internal friction angle (φ) is the angle between the envelope curve and the σ axis.

Using this approach, researchers obtain cohesive force (c) and internal friction angle (φ) values, as shown in Figure 8. Additionally, they employ the calculation method proposed by Liu et al. [35] to obtain these parameters. The primary failure mode of the specimen in the triaxial condition is shear failure, and according to the Coulomb criterion, shear strength can be determined by the cohesion force and internal friction angle.

Comparing the results obtained by the two methods, it's clear that the strength parameters obtained are very close. Any deviations between them can be primarily attributed to differences in data processing methods. In practice, taking the average of the two results can provide a more accurate representation.

The results of the study reveal that as the intermediate principal stress increases, the cohesive force (c) increases, and the internal friction angle (φ) decreases. This suggests that as the intermediate principal stress rises, the specimen compaction improves, requiring more energy for compression shear failure, resulting in a greater cohesive force. Additionally, the increase in the horizontal stress difference with the increase in intermediate principal stress leads to an increase in the force of particle interaction, contributing to a decrease in the internal friction angle.

It's important to note that due to the limited number of specimens in this test, the precise trend of changes in cohesion and internal friction angle is challenging to determine conclusively. Further research and investigation are warranted to provide a more comprehensive understanding of these relationships.

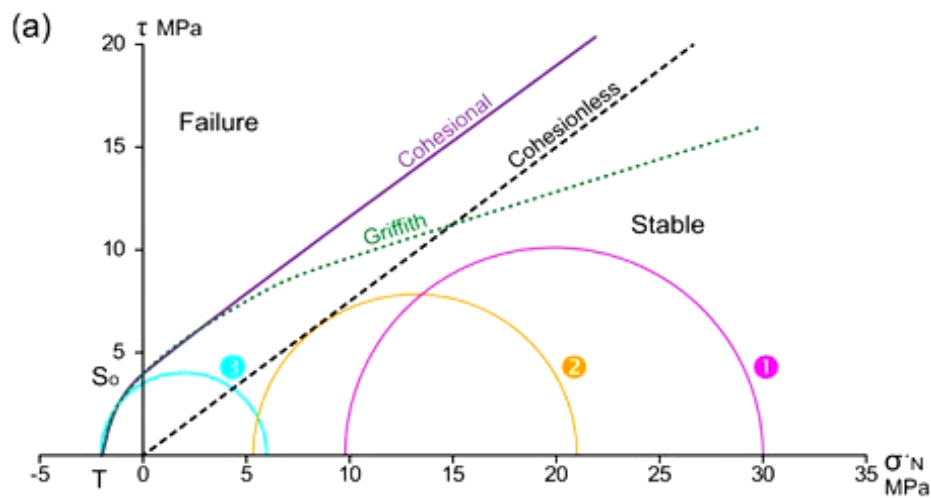

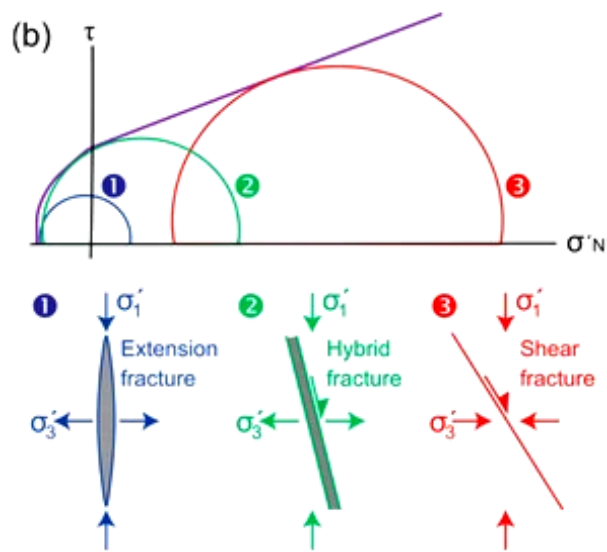

**FIGURE 8. Mohr stress circle with the same minimum and different intermediate principal stress**

The failure modes of cement specimens under true triaxial compression conditions are inferred based on variations in stress and strain, as direct observation of surface cracks and deformation characteristics is challenging due to the confinement of the true triaxial pressure chamber. The point at which the maximum principal stress approaches and exceeds the compressive strength of the cement specimens signifies their failure. Typically, there is a sharp decrease in the maximum principal stress when this occurs, indicating specimen failure. In conventional triaxial stress conditions, the typical failure modes of cylindrical specimens are illustrated in Figure 9 (uniaxial test). The failure mode appears relatively simple, dominated by shear failure, with the angle of the fracture surface related to the confining pressure. When the confining pressure is low, the fracture surface tends to be located at the top of the specimen. As the confining pressure increases, the fracture surface moves towards the middle of the specimen.

For cube specimens subjected to true triaxial stress conditions, the shear mode can be categorized into three types: V, W, and Z. V-type failures are the most common, followed by W-type, and Z-type failures are less frequent. Figure 10 illustrates the typical failure modes of cube specimens under true triaxial stress conditions. While different specimens exhibit different failure modes, the fracture surfaces are consistently parallel to the σ2 direction and form a certain angle with σ3 (ranging from 50° to 62°).

In the sixth section of the study, cohesive force and internal friction angle values for specimens under different intermediate principal stresses are obtained using two methods. According to the Coulomb criterion, a relationship between the breaking angle and the internal friction angle can be established. By applying the Coulomb criterion, the breaking angle for various intermediate principal stresses can be calculated. It is observed that the breaking angle tends to decrease with increasing intermediate principal stress.

Given that the loading range for σ2 in the test is 1232MPa, it can be predicted that the calculation range for the breaking angle is 45°65°. Upon comparing these calculated angles with the actual measured values, it is noted that the calculated range is slightly larger than the measured range. Consequently, it is evident that the relationship between the breaking angle (θ) and the internal friction angle (φ) in the Coulomb criterion has limitations when applied to the complex stress conditions experienced by rock specimens. These findings emphasize the complexities involved in characterizing failure modes and relationships between angles and stress parameters under true triaxial compression conditions.

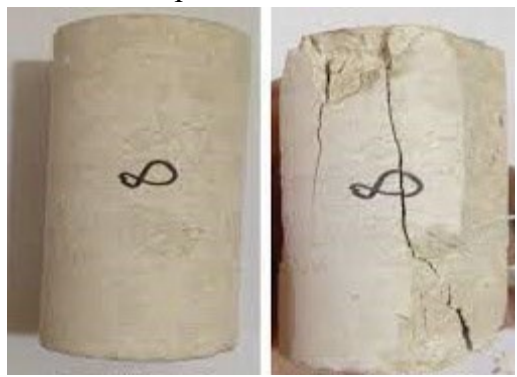

**FIGURE 9. Typical failure modes of cylindrical specimen in conventional triaxial test**

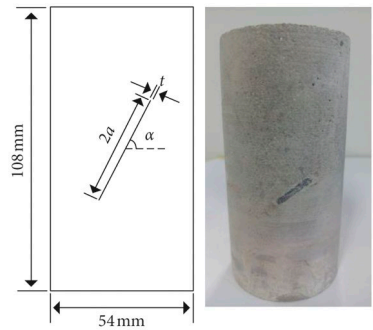

**FIGURE 10. Typical failure modes of specimen in true triaxial test**

**CONCLUSIONS**

This study presented a detailed investigation using the newly developed YSZS-2000 type multi-functional electric-hydraulic servo-controlled rock true triaxial machine, conducting true triaxial tests on 4-series similar materials. The main conclusions of the study can be summarized as follows:

Strain Behavior: The peak strain decreases as the intermediate principal stress increases for a given minimum principal stress (σ3). However, the rate of decrease in peak strain lessens as the minimum principal stress increases, eventually reaching a peak strain similar to that observed under uniaxial compression within the brittle failure range. Additionally, the relationship between Young's modulus and the confining pressure of the specimen is a monotonically increasing exponential function. Young's modulus increases with increasing intermediate principal stress for a given minimum principal stress (σ3).

Strength Parameters: Cohesive force and internal friction angle under different intermediate principal stresses were determined using two methods. It was observed that the cohesive force increases, while the internal friction angle decreases with an increase in intermediate principal stress. However, the precise behavior of cohesion and the internal friction angle remains complex and requires further investigation.

Peak Strength: In the true triaxial stress condition, peak strength increases with the increase of intermediate principal stress for a given minimum principal stress. Similarly, peak strength increases with the increase of minimum principal stress for a given intermediate principal stress. However, the rate of increase in peak strength is more pronounced with an increase in minimum principal stress compared to an increase in intermediate principal stress.

Failure Modes: In conventional triaxial stress conditions, the typical failure mode of cylindrical specimens is dominated by shear failure. The fracture surface tends to move from the top to the middle of the specimen as the confining pressure increases. For cement specimens subjected to true triaxial stress conditions, the shear mode can be classified into three types (V, W, and Z). Regardless of the specific mode, the failure surfaces in each case are parallel to the σ2 direction and exhibit a certain angle with σ3.

These conclusions provide valuable insights into the mechanical behavior and failure characteristics of materials under true triaxial stress conditions. Further research may be necessary to explore the complex relationships between various stress parameters and the detailed behavior of cohesion and the internal friction angle.